\global\let\ifmypprint\iffalse
\def\mypprint{\global\let\ifmypprint\iftrue}
\global\let\iftorefs\iffalse
\def\torefs{\global\let\iftorefs\iftrue}
\global\let\dofloatfig\iffalse
\def\floatthefig{\let\dofloatfig\iftrue}
\mypprint
\ifmypprint
\documentstyle[draft,twocolumn,pre,aps]{revtex}
\floatthefig
\else

  \iftorefs
    \catcode`\@=11
    
    \def\figure{\let\@capwidth\columnwidth\@float{figure}}
    \let\endfigure\end@float
    \@namedef{figure*}{\let\@capwidth\textwidth\@dblfloat{figure}}
    \@namedef{endfigure*}{\end@dblfloat}
    \catcode`\@=12
     \floatthefig
  \fi
\fi

\input epsf.tex

\tighten
\begin{document}
\twocolumn[\hsize\textwidth\columnwidth\hsize\csname @twocolumnfalse\endcsname

\title{Elastohydrodynamic study of actin filaments \\using
fluorescence microscopy
}
\author{D. Riveline$^1$,
C. H. Wiggins$^{2}$,
R. E. Goldstein$^{3}$, and
A. Ott$^{1,\#}$}
\address{$^1$Laboratoire PhysicoChimie Curie$^*$, Section de
Recherche, Institut Curie, \\11 rue Pierre et Marie Curie, 75231 Paris Cedex 05, France}
\address{$^2$Department of Physics, Joseph Henry Laboratories,
Princeton University, Princeton, NJ 08544}
\address{$^3$Department of Physics and Program in Applied Mathematics,
\\University of Arizona, Tucson, AZ 85721}

\date{\today}
\maketitle
\begin{abstract}
We probed the bending of actin subject to external forcing
and viscous drag. Single actin filaments were moved
perpendicular to their long axis in an oscillatory way by
means of an optically tweezed latex bead attached to one
end of the filaments. Shapes of these polymers were
observed by epifluorescence microscopy. They were found to
be in agreement with predictions of semiflexible polymer
theory and slender-body hydrodynamics. 
A persistence length of $7.4 \pm 0.2\mu$m could be
extracted.
\end{abstract}
\pacs{PACS numbers: 87.10.+e, 83.10.Nn, 42.62.-b}
]
\newcommand{\beq}{\begin{equation}}
\newcommand{\eeq}{\end{equation}}
\newcommand{\el}{\ell(\nu)}
\newcommand{\mm}{\mu{\rm m}}

In the last few years, there has been a considerable
interest in mechanical properties of actin filaments (F-
actin) and microtubules \cite{r1,r2,r3,r4,r5,r6,r7,r8,r9,r10,r11,r12,r13}. As part of the
cytoskeleton, these biological polymers play a major role
in defining shape and viscoelastic characteristics of
living cells. The corresponding proteins can be purified
and repolymerized in vitro to reach lengths of tens of microns.
The key parameter for the mechanical description
of polymers is their transverse bending modulus, often
expressed in terms of persistence length. For synthetic
polymers this parameter is generally measured by scattering
techniques  \cite{r14}. In the case of F-actin, quasi-elastic
light scattering at dilute to semi-dilute concentrations
led to a measurement of a persistence length around 0.5 $\mm$ \cite{r1,r2}, 4.5 $\mm$ \cite{r4},
or 7.5 $\mm$ \cite{r5}. End-to-end measurements by electron
microscopy gave persistence lengths of 4 $\mm$  \cite{r4}. Rhodamine
phalloidin keeps F-actin from depolymerizing, and because of
the high stiffness of biological polymers, it was possible
to measure their persistence length differently, directly
on individual filaments observed with an optical microscope
 \cite{r6,r7}. The analysis of Brownian fluctuations of actin
filaments led to various results. End-to-end measurements
gave persistence lengths of 15 $\mm$  \cite{r7}; analysis of
correlations along the filament in real space \cite{r8,r9} and a
mode analysis \cite{r10} led to about 17 $\mm$. A qualitatively
different result however was presented by K\"as et al. \cite{r11}
who, also using a mode analysis, found a wave vector
dependent persistence length in the range of 5 $\mm$ to 0.5
$\mm$.

\begin{figure} \epsfxsize=2.8 truein
\centerline{\epsffile{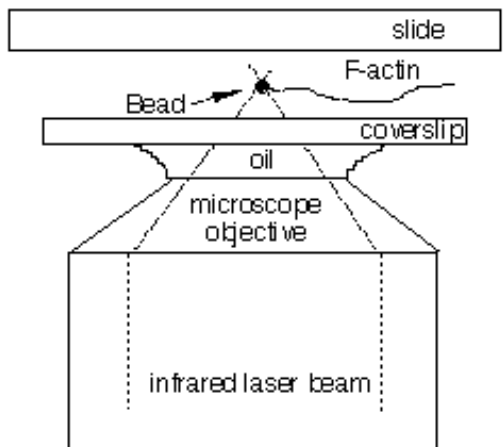}} \smallskip
\caption[]{Schematic of the observation chamber. A bead is trapped
and moved back and forth sinusoidally by optical tweezers. 
The motion of a fluorescently labeled actin filament attached
to the bead is observed by epifluorescence microscopy.}
\label{fig1}
\end{figure}

Actin filaments can be manipulated under an optical
microscope by optical tweezers via attached beads \cite{r15,r16,r17,r18}.
We present such an experiment which probes the dynamics of
individual actin filaments in order to study their bending
properties under external forcing. Similar experiments have
recently been performed for microtubules \cite{r12,r13}. A single
bead-ended filament was optically trapped (Fig. \ref{fig1}). It was
moved perpendicular to its long axis in an oscillatory way
at given frequencies such that a transverse motion was
imposed. In the case of the motion taking place in the
focal plane, its shape could be observed and then compared
to the predictions of semiflexible polymer theory and
slender-body hydrodynamics. A complete treatment of this
particular problem is given in Ref.  \cite{r19}. In the following we
recall a few results from this reference.

\dofloatfig  \fi

 A characteristic length can be extracted from the fact
that bending forces balance friction forces. Bending forces
are characterized by $A$, the elastic constant of F-actin, and
friction forces by $\zeta$, the viscous drag coefficient. The
natural length scale obtained from $A$, $\zeta$, and the frequency
of oscillation $\nu$ is
$$
\el=(A/\omega\zeta )^{1/4}= (k_BT L_p/\omega\zeta)^{1/4},
~~~{\rm with}~~\omega=2\pi\nu\nonumber
$$
or
\beq
\el=l_1\nu^{-1/4},
~~~{\rm with}~~l_1\equiv(k_BT L_p/ 2\pi \zeta)^{1/4}\label{eq2}.
\eeq
 ($k_B$ is the Boltzmann constant, $T$ the temperature, and $L_p$
 the persistence length).
This length corresponds to a longitudinal decay length of
the oscillation. We can estimate a typical value of $\el$ as
follows. The expression for $\zeta$, assuming we may locally
approximate actin as a cylinder  \cite{r20} (an additional 1/2 in
the denominator is neglected) is
\beq
 \zeta= 4\pi\eta / \ln(L/b),
\eeq
where $b$, the actin diameter, is equal to 8 nm, $L$ is the
total filament length, typically 10 $\mm$, and the viscosity of
water $\eta$ is $10^{-3}$ Pl. By taking $k_BT=4.14\cdot10^{-21} J$
 at $T=300K$,
and supposing a persistence length of 10 $\mm$ we obtain
\beq
 \el\simeq 1.39 \nu^{-1/4}\mm{\rm Hz}^{1/4}.
\eeq
The expression for the time dependent shape solves a
rescaled equation of motion  \cite{r19} derived by balancing
elastic and viscous forces and is a function of only one
free parameter, $\el$.

\dofloatfig \begin{figure} \epsfxsize=3.3 truein
\centerline{\epsffile{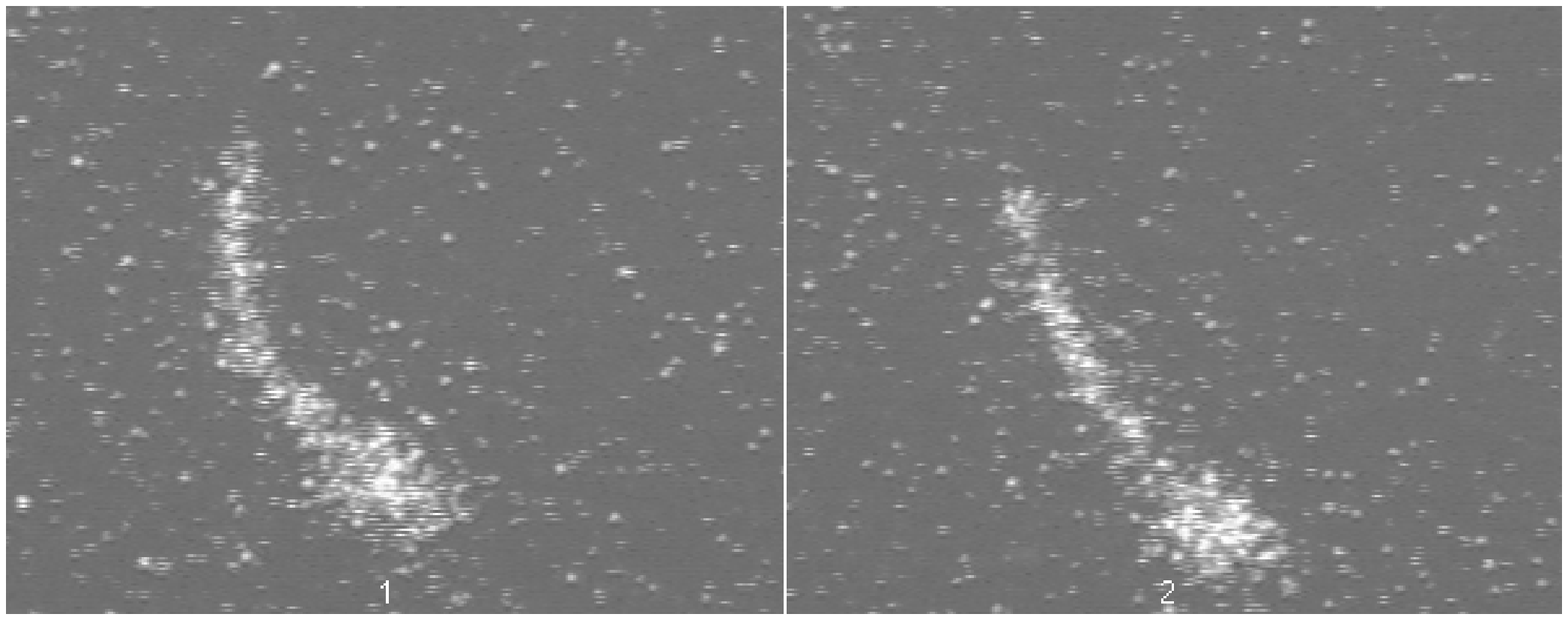}} \smallskip
\end{figure} \fi
\dofloatfig \begin{figure} \epsfxsize=3.3 truein
\centerline{\epsffile{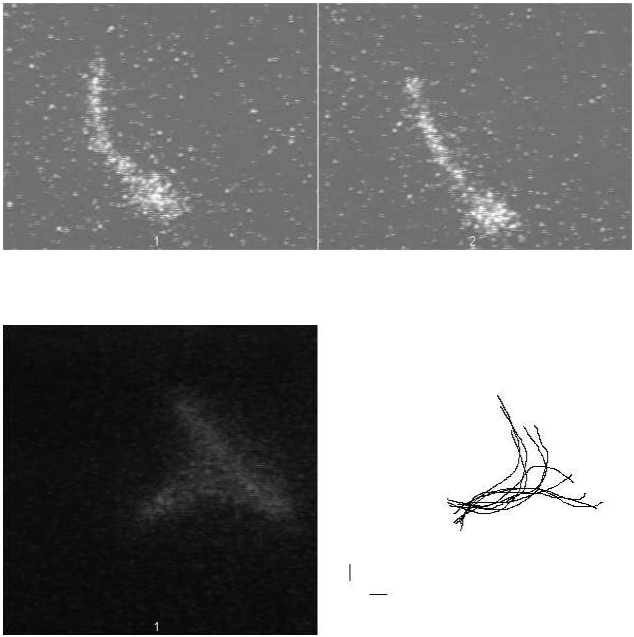}} \smallskip
\caption[]{a) Actin filament attached to a trapped bead. Time
between the images is 80 ms, and the diameter of the bead is 1 $\mm$.
b) Superposition of actin filament shapes for successive
images. Shapes were superposed and compared to theoretical
predictions only when the figure showed an axis of symmetry.
Bar is 1 $\mm$ long.}
\label{fig2}
\end{figure} \fi

 Actin was purified from chicken breast following a
published procedure. F-actin was fluorescently labeled  \cite{r6}
with rhodamine-phalloidin (R-415, Molecular Probes, Inc.).
Anti-actin antibodies (A 2668, Sigma) were covalently
coupled through carbodiimide to fluorescent polystyrene 1
$\mm$ diameter beads (CX, Duke Scientific Corporation),
following a recommended protocol (Polyscience). Experiments
were performed in actin suspension buffer containing 25 mM
imidazole, 25 mM KCl, and 5 mM $\beta$-mercaptoethanol at pH 7.65 at
room temperature. To avoid photobleaching during
observation, 1 mg/ml glucose, 33 units/ml glucose oxidase,
and 50 units/ml catalase were added to the suspension buffer.
In order to prevent actin filaments from sticking to the
glass surfaces, slides were coated with bovine serum
albumin (BSA). Observation chambers were filled by
capillarity and sealed with nail polish, the resulting cell
being around 20 $\mm$ thick. Beads and F-actin were added
separately to the observation chamber; the ratio beads:F-
actin was adjusted to have one or two beads per filament.
Among these filaments, we chose bead-ended actin filaments
for the experiment. About 30 individual filaments were
studied.

Observations were made on an inverted microscope (Axiovert
135, Zeiss) equipped with a 100 W mercury lamp, and a
standard filter set (XF37, Omega Optical). To prolong
observation time, the excitation light intensity was
reduced by inserting neutral density filters. Fluorescent
images were taken through a 4X TV tube, via an image
intensifier (C2400, Hamamatsu) followed by a CCD camera (XC-
77, Sony). Images were recorded with an S-VHS recorder. The
optical tweezer was made by focusing a 0.5 W Nd:YAG (Model
7000, Spectra-Physics) laser beam through a beam expander
and a 63X, 1.4 numerical aperture Plan-Apochromat
microscope objective (Zeiss). A mirror mounted on a
galvanometer (6350, Cambridge Technology) connected to a
function generator served to trap and move beads in the
focal plane (Fig. \ref{fig1}). The trap was driven by sine waves,
with frequencies and amplitudes ranging from
0.1 Hz to 6 Hz and from 5 $\mm$ to 10 $\mm$, respectively. Below
0.1 Hz the motion of actin appeared forceless, and above 6
Hz acquisition was limited by the video rate. Up to 10
images per oscillation period were digitized with a Power
Macintosh equipped with a Scion frame grabber card; they
were subsequently analyzed using NIH-Image software. Figure
\ref{fig2}a shows a typical image. Images corresponded to an overall
screen size of 44 $\mm$ in horizontal dimension leading to a
value of 0.06 $\mm$/pixel after digitization.
To determine the coordinates of the filament, the shape
of each image was drawn with the mouse starting
from the center of the bead. 
During the experiment out-of-focus movements of
the filament could bias the shapes; a flow could not always
be avoided. In order to detect these artifacts, successive
shapes were superposed (Fig. \ref{fig2}b). Only those sequences that
met the two following criteria were retained for further
analysis: i) filaments remained in the focal plane for at
least an oscillation period; and ii) the axis of
symmetry was clearly visible upon superposition of subsequent
images.

\begin{figure} \epsfxsize=3.3 truein
\centerline{\epsffile{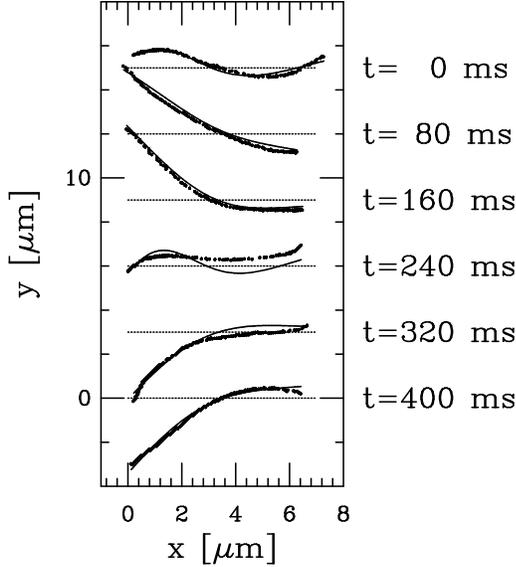}} \smallskip
\caption[]{Successive filament shapes (circles) at 2 Hz
with the corresponding fit (continuous lines) following the
theoretical prediction (see Ref.  \cite{r19}).}
\label{fig3}
\end{figure}

The expression for the shape $y(x,t)$ solves the equation
of motion and is a function of $x/\el$ and $L/\el$  \cite{r19}.
Since the amplitude and phase of the motion could be read directly from
the position of the driving bead, $\el$ was the only free
parameter. 
We minimized $\chi^2$ for each of
the 221 selected images. Figure \ref{fig3} shows a typical fit for a
half period of oscillation.
\dofloatfig  \fi
\dofloatfig \begin{figure} \epsfxsize=3.3 truein
\centerline{\epsffile{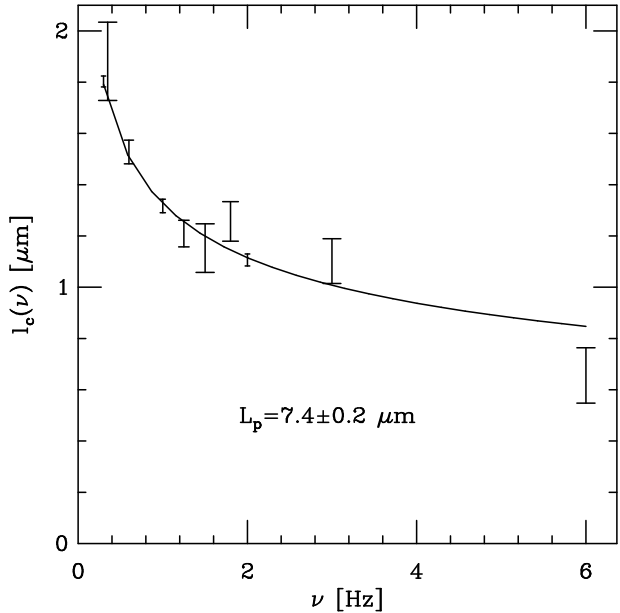}} \smallskip
\caption[]{The characteristic length scale $\el$ as a function of
driving frequency $\nu$.}
\label{fig4}
\end{figure} \fi
The value of $\el$ was extracted from each image.
Then, fitting the relation $\el$ to the expression
$\el=l_1\nu^{-\alpha}$, we found the values of $l_1$ and $\alpha$.
We found $\alpha=.26\pm.04$, indicating
good agreement with the prediction $\alpha=1/4$ (see Eq. \ref{eq2}).
As defined above, $l_1$ is a function of known quantities ($k_BT,\zeta$)
and one unknown, $L_p$. Enforcing $\alpha=1/4$, we are left with a 
$1$-parameter fit, from which we determined $L_p=7.4\pm0.2 \mm$
(see Fig. \ref{fig4}).
Histograms of the different $l(\nu)$ for a given
$\nu$ show that the data are not always distributed
symmetrically about the mean; a few short $l(\nu)$
values increase the distribution at one end. By
fitting with and without these outliers, however, we showed they
do not significantly affect the results.

 As shown in figure \ref{fig4}, our experimental data are in agreement with
the given scaling law (Eq. \ref{eq2}). Even in the case where modes are
imposed by external forcing, dynamics of an individual
actin filament follow the theory of semiflexible polymers.
Previous authors found the persistence length
of rhodamine F-actin to be $9 \mm$ as opposed to $18 \mm$ for
rhodamine phalloidin actin  \cite{r9}. Compared to
other measurements on rhodamine phalloidin
labeled F-actin \cite{r7,r8,r9,r10,r11} our value is higher than those
presented in Ref.  \cite{r11}, and appears lower than those
presented in Ref.  \cite{r7,r8,r9,r10}. A lower persistence length would
suggest that actin may be weakened by the external forcing.
Further experiments using different techniques would be
needed to confirm this hypothesis. 

To avoid longitudinal
tension in our experiment, smaller amplitude oscillations
should be applied; the linearized analysis is more
precise and appropriate in this case. However optical
resolution, the size of the trapped bead, and camera
sensitivity impose close limits.
In our experiment, errors are due to several causes, only
some of which could in principle be eliminated. The
parameter $\el$ extracted from our observations is of the
order of a micron. The values measured therefore not
only approach the limit of
optical resolution, but also the size of the driving bead. 
The hydrodynamics
is simplified and leaves out the flow due to the bead,
which is a limitation of the analysis. 
Brownian motion of the filament, giving some more disordered
motion than that predicted by theory, increases the
error at low frequencies. Furthermore the
persistence length depends on the 4th root of the
key
parameter $\el$. 
In future experiments, attention should be paid
to an external measurement of the phase of the sine-motion which
simplifies the data analysis.
For these reasons our systematic error of the persistence length
is higher than those of previous publications, but keeping in
mind the scattering of the results presented by different authors using the same or different techniques,
this does not appear as a significant drawback. 

We have shown that dynamic behaviour of an individual actin
filament can be predicted by describing F-actin as a
semiflexible polymer. This experiment constitutes a new way
for studying elasticity of semiflexible biopolymers. The
technique could be applied to the study of F-actin and
microtubule changes in elasticity, when they interact with
associated proteins.


We are grateful to Olivier Cardoso for his modification of
NIH Image Software, and Olivier Cardoso, Olivier Thoumine,
Frank J\"ulicher and Jacques Prost for stimulating
discussions. This work benefited from a grant from the
Institut Curie.
This work was supported by NSF PFF Grant DMR 93-50227 and the
A.P. Sloan Foundation (REG).


\begin{references}

\bibitem[\#]{ott}To whom correspondence should be addressed
\bibitem[*]{curie}Laboratoire associ\'e au Centre 
National de la Recherche Scientifique (CNRS) et \`a l'Universit\'e Paris 6.
\bibitem{r1} C.F. Schmidt, M. B\"armann, G. Isenberg, and E. Sackmann,
Macromolecules {\bf 22}, 3638 (1989).
\bibitem{r2} T. Piekenbrock and E. Sackmann, Biopolymers {\bf 32}, 1471
 (1992).
\bibitem{r3} E. Farge and A. C. Maggs, Macromolecules {\bf 26}, 5041
(1993).
\bibitem{r4} T. Takebayashi, Y. Morita, and F. Oosawa, Biochem.
Biophys. Acta {\bf 492}, 357 (1977).
\bibitem{r5} J. Dr\"ogemeier and W. Eimer, Macromolecules {\bf 27}, 96
(1994).
\bibitem{r6} J. Pardee and J. A. Spudich, Methods Enzymol. {\bf 85B}, 164
(1982).
\bibitem{r7} T. Yanagida, M. Nakase, K. Nishiyama, and F. Oosawa,
Nature {\bf 307}, 58 (1984).
\bibitem{r8} A. Ott, M. Magnasco, A. Simon, and A. Libchaber, Phys.
Rev. E {\bf 48}, 1642 (1993).
\bibitem{r9} H. Isambert, P. Venier, A. C. Maggs, A. Fattoum, R.
Kassab, D. Pantaloni, and M.-F. Carlier, J. Biol. Chem.
{\bf 270}, 11437 (1995).
\bibitem{r10} F. Gittes, B. Mickey, J. Nettleton, and J. Howard, J.
Cell Biol. {\bf 120}, 923 (1993).
\bibitem{r11} J. K\"as, H. Strey, M. B\"armann, and E. Sackmann,
Europhys. Lett. {\bf 21}, 865 (1993).
\bibitem{r12} H. Felgner, R. Frank, and M. Schliwa, J. Cell Sc. {\bf 109},
509 (1996).
\bibitem{r13} M. Kurachi, M. Hoshi, and H. Tashiro, Cell Motil.
Cytoskel. {\bf 30}, 221 (1995).
\bibitem{r14} B.J. Berne and R. Pecora, Dynamic Light Scattering (J.
Wiley, New-York, 1976).
\bibitem{r15} S. M. Block, L. S. Goldstein, and B. J. Schnapp,
Nature {\bf 348}, 348 (1990).
\bibitem{r16} J. T. Finer, R. M. Simmons, and J. A. Spudich, Nature
{\bf 368}, 113 (1994).
\bibitem{r17} T. T. Perkins, S. R. Quake, D. E. Smith, and S. Chu,
Science {\bf 264}, 822 (1994).
\bibitem{r18} N. Suzuki, H. Miyata, S. Ishiwata, and K. Kinosita
Jr., Biophys. J. {\bf 70}, 401 (1996).
\bibitem{r19} C. Wiggins, D. Riveline, A. Ott, and R. Goldstein, 
submitted to Biophysical Journal, cond-mat/9703244
\bibitem{r20} E. Guyon, J.-P. Hulin, and L. Petit, Hydrodynamique
physique, (Editions du CNRS, Meudon, 1991), p. 359.
\end{references}
\end{document}
\end